\documentclass[secnumarabic, floatfix, nofootinbib,tightenlines,nobibnotes, aps,showpacs,prd,twocolumn,preprintnumbers,amsmath,amssymb,superscriptaddress]{revtex4-1}

\usepackage{amsmath,amssymb,graphicx,color,dcolumn,bm}
\usepackage[hypertex,bookmarks=true]{hyperref}
\usepackage{setspace,atbegshi,graphicx,color}
\def\bra#1{\langle #1|}
\def\ket#1{|#1\rangle}

\newcommand{\cC}{{\mathcal{C}}}

\newcommand{\cS}{{\mathcal{S}}}
\newcommand{\cT}{{\mathcal{T}}}

\begin{document}

\title{Effective dimension, level statistics, and integrability of Sachdev-Ye-Kitaev-like models}

\author{Eiki Iyoda}
\affiliation{
Department of Applied Physics, The University of Tokyo,
7-3-1 Hongo, Bunkyo-Ku, Tokyo 113-8656, Japan
}
\author{Hosho Katsura}
\affiliation{
Department of Physics, The University of Tokyo, 
7-3-1 Hongo, Bunkyo-Ku, Tokyo 113-8656, Japan
}
\author{Takahiro Sagawa}
\affiliation{
Department of Applied Physics, The University of Tokyo,
7-3-1 Hongo, Bunkyo-Ku, Tokyo 113-8656, Japan
}

\begin{abstract}
The Sachdev-Ye-Kitaev (SYK) model attracts attention in the context of information scrambling, which represents delocalization of quantum information and is quantified by the out-of-time-ordered correlators (OTOC). The SYK model contains $N$ fermions with disordered and four-body interactions. Here, we introduce a variant of the SYK model, which we refer to as the Wishart SYK model. We investigate the Wishart SYK model for complex fermions and that for hard-core bosons. We show that the ground state of the Wishart SYK model is massively degenerate and the residual entropy is extensive, and that the Wishart SYK model for complex fermions is integrable. In addition, we numerically investigate the OTOC and level statistics of the SYK models. At late times, the OTOC of the fermionic Wishart SYK model exhibits large temporal fluctuations, in contrast with smooth scrambling in the original SYK model. We argue that the large temporal fluctuations of the OTOC are a consequence of a small effective dimension of the initial state. We also show that the level statistics of the fermionic Wishart SYK model is in agreement with the Poisson distribution, while the bosonic Wishart SYK model obeys the GUE or the GOE distribution.
\end{abstract}

\maketitle

\section{Introduction}
Scrambling of quantum information in quantum many-body systems attracts attention in a wide range of fields including high energy physics and condensed matter physics.
The Sachdev-Ye-Kitaev (SYK) model exhibits a fascinating feature of scrambling, which is a quantum model of fermions with disordered, all-to-all, and four-body interactions~\cite{KitaevKITP1,KitaevKITP2,Polchinski2016JHEP,Maldacena2016PRD94}. 
Recently, Kitaev proposed the SYK model to address the black hole information paradox~\cite{KitaevKITP1,KitaevKITP2}. 
In this context, it was conjectured that black holes are the fastest scramblers of quantum information~\cite{Hayden2007JHEP120,SekinoJHEP1126-6708-2008-10-065,Shenker2014JHEP67,Maldacena2016JHEP106,Hosur2016JHEP4}, where scrambling behavior has been investigated with the decay of the out-of-time-ordered correlators (OTOC)~\cite{KitaevKITP1,KitaevKITP2,Hosur2016JHEP4,Maldacena2016JHEP106,ALEINER2016378AnnPhys,Haehl2017qflarXiv,Roberts2017JHEP121,Caputa2016doi10.1093/ptep/ptw157,Kukuljan2017PhysRevB.96.060301,Rozenbaum2017PhysRevLett.118.086801,Hashimoto2017JHEP138} and the negativity of tripartite mutual information (TMI)~\cite{Hosur2016JHEP4,CERF199862PhysicaD}.
Taking advantage of the fact that the SYK model is tractable, i.e., the two-point and four-point functions can be calculated analytically in the limit of large-$N$ and large disorder (or low energy)~\cite{Polchinski2016JHEP,Maldacena2016PRD94}, it is shown that the SYK model exhibits the fastest scrambling and saturates the upper bound of the decay rate of the OTOC (``the bound on chaos")~\cite{Maldacena2016JHEP106,Polchinski2016JHEP,Maldacena2016PRD94}. 

While the SYK model was originally introduced with Majorana fermions~\cite{KitaevKITP1,KitaevKITP2}, the SYK model with complex fermions~\cite{Sachdev2015PhysRevX.5.041025,Fu2016PhysRevB.94.035135} can also be defined as
\begin{align}
\label{eq:SYKmodel}
H_\mathrm{SYK}
:=
\frac{1}{N^{3/2}}
\sum_{1\leq j<i\leq N \atop 1\leq k<l\leq N}
J_{i,j;k,l}
c_i^\dag c_j^\dag
c_k c_l,
\end{align}
where $N$ is the number of sites, and $c_i$ ($c_i^\dag$) is the annihilation (creation) operator of a complex fermion at site $i$, satisfying the anti-commutation relation $\{c_i,c_j^\dag\}=\delta_{i,j}$ and $\{c_i,c_j\}=\{c_i^\dag,c_j^\dag\}=0$. The coupling constant $J_{i,j;k,l}$ is sampled from the complex Gaussian distribution with variance $J^2$, satisfying $J_{i,j;k,l}=J_{l,k;j,i}^*$.  
Besides, other variants of the SYK model have been investigated not only in high energy physics, but also in condensed matter physics~\cite{Sachdev2015PhysRevX.5.041025,Fu2016PhysRevB.94.035135,Danshita2017doi:10.1093/ptep/ptx108,PikulinPhysRevX.7.031006,Franz1808.00541,Song2017PhysRevLett.119.216601,Jian2017PhysRevLett.119.206602,Gu2017JHEP125,Chen2017PhysRevLett.119.207603,Bi2017PhysRevB.95.205105,You2017PhysRevB.95.115150,Chen2017JHEP150,Garcia2018PhysRevLett.120.241603,Zhang2018PhysRevB.97.201112} because of its relevance to non-Fermi liquid, the quantum critical phenomena, and the effect of disorder in strongly correlated systems. For example, there are many extensions of the SYK model: that with $q$-point interactions~\cite{Maldacena2016PRD94}, that with hard-core bosons~\cite{Fu2016PhysRevB.94.035135}, SUSY extensions~\cite{Fu2017PhysRevD.95.026009,Sannomiya2017PhysRevD.95.065001,Kanazawa2017JHEP50,Peng2017JHEP62,Li2017JHEP111,Garcia2018_PhysRevD.97.106003}, disorder-free tensor models~\cite{Witten2016iuxarXiv,Peng2017JHEP62}, the SYK model with a lattice structure~\cite{Gu2017JHEP125,Berkooz2017JHEP138,Jian2017PhysRevLett.119.206602,Chowdhury1801.06178,Patel2018_PhysRevX.8.021049,Haldar2018_PhysRevB.97.241106}, 
and a kind of coupled or perturbed system~\cite{Song2017PhysRevLett.119.216601,Bi2017PhysRevB.95.205105,Chen2017PhysRevLett.119.207603,Chen2017JHEP150,Garcia2018PhysRevLett.120.241603,Zhang2018PhysRevB.97.201112}. 
Furthermore, experimental implementation of the SYK model has been theoretically proposed with ultracold atoms and solid-state devices~\cite{Danshita2017doi:10.1093/ptep/ptx108,PikulinPhysRevX.7.031006,Franz1808.00541}.

In this paper, we investigate a variant of the SYK model, which we refer to as the Wishart SYK model.
This model reduces to a clean SYK model without quenched disorder as a special case.
In a previous work~\cite{Iyoda2018PhysRevA.97.042330}, two of the authors found that the clean SYK model exhibits large temporal fluctuations in contrast to the original SYK model. Here we will investigate the origin of such large fluctuations from a more general perspective based on the Wishart SYK model.

We find that the ground state of the Wishart SYK model is very degenerate, and the residual entropy is extensive. The degeneracy makes the effective dimension of the initial state smaller, and the small effective dimension leads to large temporal fluctuations. On the other hand, the original SYK model shows a large effective dimension and small temporal fluctuations. 

We also numerically investigate the level statistics of the original and the Wishart SYK models. We show that the level statistics of the fermionic Wishart SYK model is in good agreement with the Poisson distribution. Correspondingly, we prove that the fermionic Wishart SYK model is integrable by mapping it onto a particular case of the Richardson-Gaudin model~\cite{Richardson1965doi:10.1063/1.1704367,gaudin_2014}, which is known to be Bethe-ansatz solvable.

The rest of this paper is organized as follows. In Sec.~\ref{sec:model}, we introduce the Wishart SYK model and investigate its basic properties. In Sec.~\ref{sec:energy}, we show the numerical results of the energy spectrum of the original and the Wishart SYK models. In Sec.~\ref{sec:OTOC}, we numerically show dynamics of the OTOC and the effective dimension. In Sec.~\ref{sec:level}, we investigate the level statistics of the energy level spacings. In Sec.~\ref{sec:integrability}, we show the integrability of the fermionic Wishart SYK model. 
In Appendix~\ref{sec:AppEqualities}, we show that the equalities of (\ref{eq:NumZES0}) and (\ref{eq:NumZES}) hold for the fermionic model.
In Appendix~\ref{sec:AppAntiUnitary}, we construct an anti-unitary operator commuting with the Hamiltonian of the Wishart SYK model. 
In Appendix~\ref{sec:AppIndependence}, we show the linear independence of the mutually commuting operators of the fermionic Wishart SYK model.
In Appendices~\ref{sec:AppNotation} and \ref{sec:AppABA}, we review some previous results on the symmetry algebra and the algebraic Bethe ansatz for the Richardson-Gaudin model. 

\section{Wishart SYK model}
\label{sec:model}

\subsection{Hamiltonian}
\label{sec:modelHam}
We first define the Wishart SYK model, which is named after the Wishart matrices in random matrix theory. The Hamiltonian of the Wishart SYK model is defined as
\begin{align}
\label{eq:Ham_wSYK}
H_{\mathrm{wSYK}}
&:=
Q^\dag Q,
\\
\label{eq:Q_wSYK}
Q
&:=
\frac{1}{N}\sum_{1\leq k<l\leq N} J_{k,l}c_k c_l,
\end{align}
where $c_k$ is the annihilation operator of a complex fermion, and the coupling constant $J_{k,l}$ is sampled from the complex Gaussian distribution with mean $J_\mathrm{mean}$ and variance $J^2$.
The total fermion number $N_{\mathrm{P}}:=\sum_{i=1}^N c_i^\dag c_i$ is conserved: $[H_{\mathrm{wSYK}},N_{\mathrm{P}}]=0$.

The Wishart SYK model includes a clean counterpart of the SYK model as a special case, where the coupling constant $J_{i,j;k,l}$ is uniform in Eq.~(\ref{eq:SYKmodel}).
This limit is achieved by setting $J=0$ and $J_\mathrm{mean}\neq 0$ in Eqs.~(\ref{eq:Ham_wSYK}) and (\ref{eq:Q_wSYK}).

\subsection{Ground-state degeneracy}
\label{sec:modelZES}
Since the Hamiltonian of the Wishart SYK model is positive-semidefinite, if there are eigenstates whose energies are zero, they are ground states. In the following way, we find a huge number of the ground states in the Wishart SYK model. 

The operator $Q$ annihilates two fermions and decreases the total fermion number by $2$. When $Q$ acts on a sector of the total fermion number $N_{\mathrm{P}}$, the change in the dimension of the sector is 
\begin{align}
C(N,N_{\mathrm{P}}-2)-C(N,N_{\mathrm{P}}),
\end{align}
where $C(n,k)$ is the binomial coefficient.
When $N_{\mathrm{P}}=0$ or $1$, the second argument of $C(N,N_{\mathrm{P}}-2)$ can be negative. In such cases, $C(N,N_{\mathrm{P}}-2)$ is regarded as $0$.
The change is negative when $N_{\mathrm{P}}\leq \lfloor \frac{N+1}{2}\rfloor$, where $\lfloor \cdot \rfloor$ is the floor function.
If the kernel of the operator $Q$ restricted to the sector is not null, there exist eigenstates of $H_{\mathrm{wSYK}}$ whose eigenvalues are zero. Thus, there are zero-energy eigenstates when $N_{\mathrm{P}}\leq \lfloor \frac{N+1}{2}\rfloor$.

We denote by $Z_{N,N_{\mathrm{P}}}$ the number of the zero-energy states in the sector of the fermion number $N_{\mathrm{P}}$.
To estimate a lower bound of $Z_{N,N_{\mathrm{P}}}$, we apply the rank-nullity theorem, which is given by
\begin{align}
\label{eq:RankNullity}
\mathrm{dim}[\mathrm{Im}(Q)]+\mathrm{dim}[\mathrm{Ker}(Q)]=\mathrm{dim}[\mathrm{Domain}(Q)],
\end{align} 
where $\mathrm{Im}$, $\mathrm{Ker}$, and $\mathrm{Domain}$ respectively represent the image, the kernel, and the domain of an operator.
In the sector of the fermion number $N_\mathrm{P}$,
$\mathrm{dim}[\mathrm{Ker}(Q)]=Z_{N,N_{\mathrm{P}}}$,
$\mathrm{dim}[\mathrm{Domain}(Q)]=C(N,N_{\mathrm{P}})$, and $\mathrm{dim}[\mathrm{Im}(Q)]\leq C(N,N_{\mathrm{P}}-2)$ hold, where the equality in the last inequality is achieved if $Q$ is surjective.
Thus, a lower bound of $Z_{N,N_{\mathrm{P}}}$ with $N_{\mathrm{P}}\leq \lfloor \frac{N+1}{2}\rfloor$ is given by
\begin{align}
\label{eq:NumZES0}
Z_{N,N_{\mathrm{P}}}
\geq
C(N,N_{\mathrm{P}})-C(N,N_{\mathrm{P}}-2).
\end{align}
Defining $Z_N:=\sum_{N_{\mathrm{P}}=0}^{N}Z_{N,N_{\mathrm{P}}}$ and using inequality~(\ref{eq:NumZES0}), we obtain a lower bound of $Z_N$ as
\begin{align}
\nonumber
Z_N
&\geq
\sum_{N_{\mathrm{P}}=0}^{\lfloor \frac{N+1}{2}\rfloor}
Z_{N,N_{\mathrm{P}}}
\\
\nonumber
&=
\sum_{N_{\mathrm{P}}\geq 1, N_{\mathrm{P}}\in\mathrm{odd}}^{\lfloor \frac{N+1}{2}\rfloor}
Z_{N,N_{\mathrm{P}}}
+
\sum_{N_{\mathrm{P}}\geq 0, N_{\mathrm{P}}\in\mathrm{even}}^{\lfloor \frac{N+1}{2}\rfloor}
Z_{N,N_{\mathrm{P}}}
\\
\nonumber
&\geq
C\left(N,\left\lfloor\frac{N+1}{2}\right\rfloor\right)
+
C\left(N,\left\lfloor\frac{N+1}{2}\right\rfloor-1\right)
\\
\label{eq:NumZES}
&=
C\left(N+1,\left\lfloor\frac{N+1}{2}\right\rfloor\right).
\end{align}
In Appendix~\ref{sec:AppEqualities}, we show that the equalities of (\ref{eq:NumZES0}) and (\ref{eq:NumZES}) indeed hold for the Hamiltonian (\ref{eq:Ham_wSYK}), where the counting of the zero-energy states arrives at that of the lowest weight states of the total angular momentum.

Since $Z_N$ increases exponentially with $N$, the Wishart SYK model has an extensive residual entropy.
This fact reminds us of the residual entropy of the original SYK model. We should note that the residual entropy in the original SYK model does not represent huge degeneracy in the ground state but many low-energy excited states near the ground state in the large-$N$ limit~\cite{Maldacena2016PRD94}.

We also consider the SYK model and the Wishart SYK model with hard-core bosons. They are defined by replacing the annihilation (creation) operator for fermions $c_i$ ($c_i^\dag$) by that for hard-core bosons $b_i$ ($b_i^\dag$), which satisfy $[b_i,b_j^\dag]=[b^\dag_i,b_j^\dag]=0$ for $i\neq j$, 
$b_i^2={(b_i^\dag)}^2=0$ and $\{b_i,b_i^\dag\}=1$. 
In this paper, we refer to the (Wishart) SYK model for complex fermions/hard-core bosons as the fermionic/bosonic (Wishart) SYK model. In the same way as the fermionic Wishart SYK model, we show that the Wishart SYK model for hard-core bosons has the same ground-state degeneracy as the fermionic Wishart SYK model. We note that the discussion in Appendix~\ref{sec:AppEqualities} does not apply for the bosonic model. However, we numerically confirmed that the lower bounds (\ref{eq:NumZES0}) and (\ref{eq:NumZES}) are indeed saturated for the bosonic Wishart SYK model.

We remark on a dis-similarity between the Hamiltonian of the Wishart SYK model (\ref{eq:Ham_wSYK}) and the Hamiltonian of $\mathcal{N}=2$ SUSY SYK model~\cite{Fu2017PhysRevD.95.026009,Sannomiya2017PhysRevD.95.065001,Kanazawa2017JHEP50,Peng2017JHEP62,Li2017JHEP111} defined as
\begin{align}
\label{eq:Ham_SUSY_SYK}
H_{\mathrm{SUSY},\mathcal{N}=2}
&:=
\mathcal{Q}^\dag \mathcal{Q}+\mathcal{Q} \mathcal{Q}^\dag,
\\
\mathcal{Q}
&:=
\frac{i}{N}\sum_{1\leq i<j<k\leq N} J_{i,j,k}c_i c_j c_k,
\end{align}
where $J_{i,j,k}$ are independent complex Gaussian variables with variance $J^2$, and the supercharge $\mathcal{Q}$ is nilpotent: $\mathcal{Q}^2=0$.
It is shown that there are zero-energy ground states in the $\mathcal{N}=2$ SUSY SYK model because there is nonzero subspace spanned by the vectors with $\mathcal{Q}\ket{\psi}=\mathcal{Q}^\dag\ket{\psi}=0$.
When we assume that $N$ is even and $N_{\mathrm{P}}=N/2$ for simplicity, the number of the zero-energy ground states of the $\mathcal{N}=2$ SUSY SYK model is given by $2\cdot 3^{N/2-1}$ as shown in Ref.~\cite{Kanazawa2017JHEP50}, which is much smaller than $Z_{N,N/2}$ for the Wishart SYK model.
Such a SUSY extension can be defined when the supercharge is a product of $q$ annihilation operators with odd $q$. In the case of the Wishart SYK model, $Q$ is defined with two annihilation operators.
The difference between them is remarkable when we consider the fermionic parity $(-1)^{N_{\mathrm{P}}}$.
While the supercharge $\mathcal{Q}$ anti-commutes with the fermionic parity $\{\mathcal{Q},(-1)^{N_{\mathrm{P}}}\}=0$ for the SUSY case, the operator $Q$ in the fermionic Wishart SYK model commutes with the fermionic parity $[Q,(-1)^{N_{\mathrm{P}}}]=0$.

The SYK model and the SUSY SYK model have been investigated from the viewpoint of the random matrix theory.
While the spectral density of the SYK model is characterized by the Gaussian ensembles~\cite{Garcia2016_PhysRevD.94.126010,Garcia2017_PhysRevD.96.066012}, that of the SUSY SYK model is generically described by the Wishart-Laguerre ensembles~\cite{Li2017JHEP111,Garcia2018_PhysRevD.97.106003}. 
We note that why we name the model~(\ref{eq:Ham_wSYK}) after the Wishart SYK model is because the operator $Q$ is represented as a rectangular matrix, which directly leads to the huge ground-state degeneracy of the Wishart SYK model.

\section{Energy spectrum}
\label{sec:energy}

In this section, we show the results of numerically exact diagonalization of the Hamiltonian of the SYK models.
We set $J_\mathrm{mean}=0$ in the following sections.
In this section, the numerical results are obtained from a single disorder realization.

\begin{figure}[t]
\begin{center}
\includegraphics[width=\linewidth]{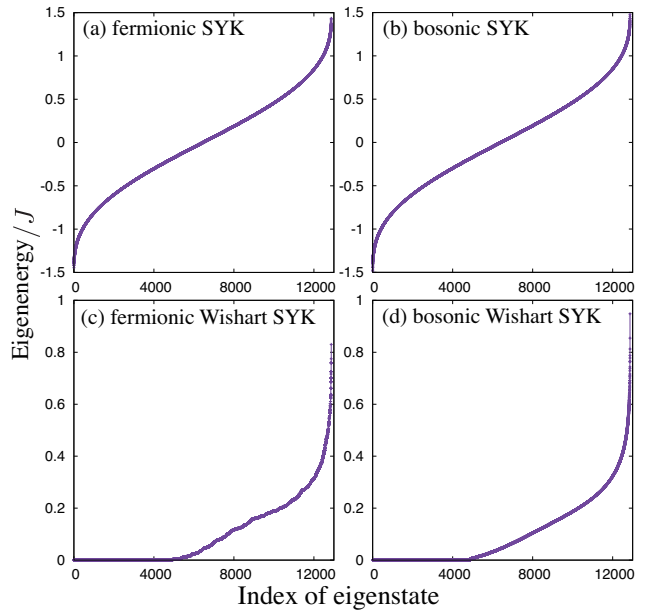}
\end{center}
\caption{
Eigenenergies of $H_\mathrm{SYK}$ and $H_\mathrm{wSYK}$ for $N=16$ and $N_{\mathrm{P}}=N/2$. 
Energy levels are labeled in ascending order from the lowest to the highest.
}
\label{fig:Energy_SYK}
\end{figure}
Figure~\ref{fig:Energy_SYK} shows eigenenergies of the SYK model with $N=16$ and $N_{\mathrm{P}}=N/2$. 
Figures~\ref{fig:Energy_SYK} (a) and (b) show that the structures of eigenenergies are very similar between the fermionic and bosonic SYK models. We note that all the eigenenergies are non-degenerate in the sector of $N_{\mathrm{P}}=N/2$. When we look at the sector of $N_{\mathrm{P}}=0$ and $1$ (not shown), there are degenerate energy eigenstates at $E=0$ whose degeneracy is $N+1$. This degeneracy at $E=0$ comes from the fact that the sectors of $N_{\mathrm{P}}=0$ and $1$ are in the kernel of the Hamiltonian of the SYK model.

Figures~\ref{fig:Energy_SYK} (c) and (d) show eigenenergies of the fermionic/bosonic Wishart SYK models. All the eigenenergies are non-negative because the Hamiltonian is positive-semidefinite. 
A huge number of the ground states are observed as discussed in Sec.~\ref{sec:model}. We have checked that the ground-state degeneracy equals the right-hand side of Eq.~(\ref{eq:NumZES0}). For example, the degeneracy with $N=16$ and $N_{\mathrm{P}}=8$ is $Z_{N,N_{\mathrm{P}}}=4862$. Concerning the excited states, the structure of the energy spectrum differs between the fermionic and the bosonic Wishart models.
While the spectrum of the bosonic Wishart model is smooth, that of the fermionic Wishart model is rough.

\begin{figure}[t]
\begin{center}
\includegraphics[width=\linewidth]{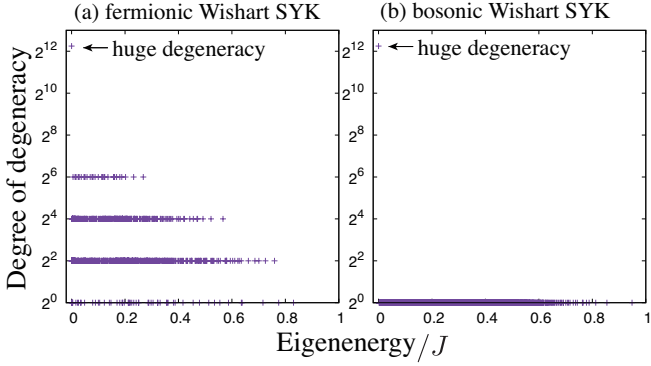}
\end{center}
\caption{
Energy dependence of the degeneracy of $H_\mathrm{wSYK}$ for $N=16$ and $N_{\mathrm{P}}=N/2$. 
}
\label{fig:Deg_wSYK}
\end{figure}

Figure~\ref{fig:Deg_wSYK} shows the degeneracy of the eigenenergies of the Wishart SYK models.
As shown in Fig.~\ref{fig:Deg_wSYK} (b), there is no degeneracy in the excited states of the bosonic Wishart SYK model.
On the other hand, Fig.~\ref{fig:Deg_wSYK} (a) shows that there are many degenerate excited states in the fermionic Wishart SYK model.
The degeneracy is given by $2^l$, where $l=0,2,4$, and $6$. 
We also note that the degeneracy tends to decrease as the energy increases. 
In general, the same degeneracy is also seen in other disorder realizations.
We will examine this degeneracy in the excited states in Sec.~\ref{sec:integrability}.

\section{Dynamics of OTOC and effective dimension}
\label{sec:OTOC}
In this section, we numerically investigate the effect of the huge ground-state degeneracy on dynamics of the Wishart SYK model.
We focus on the out-of-time-ordered correlator (OTOC), which is an indicator of scrambling.
An OTOC for operators $A$ and $B$, with an initial state $\ket{\Psi}$, and at time $t$ is defined as
\begin{align}
C_{\mathrm{AB}}(t)
:=
\bra{\Psi}A^\dag (t) B^\dag(0) A(t) B(0)\ket{\Psi}.
\end{align}
We also define the long time average of the OTOC and the temporal fluctuations of its real part.
\begin{align}
\overline{C_{\mathrm{AB}}(t)}
&:=
\frac{1}{T}\int_0^T dt C_{\mathrm{AB}}(t),
\\
(\Delta C_{\mathrm{AB}})^2
&:=
\overline{
(
\mathrm{Re}C_{\mathrm{AB}}(t)
-
\mathrm{Re}\overline{C_{\mathrm{AB}}(t)}
)^2
}.
\end{align}

\begin{figure}[t]
\begin{center}
\includegraphics[width=\linewidth]{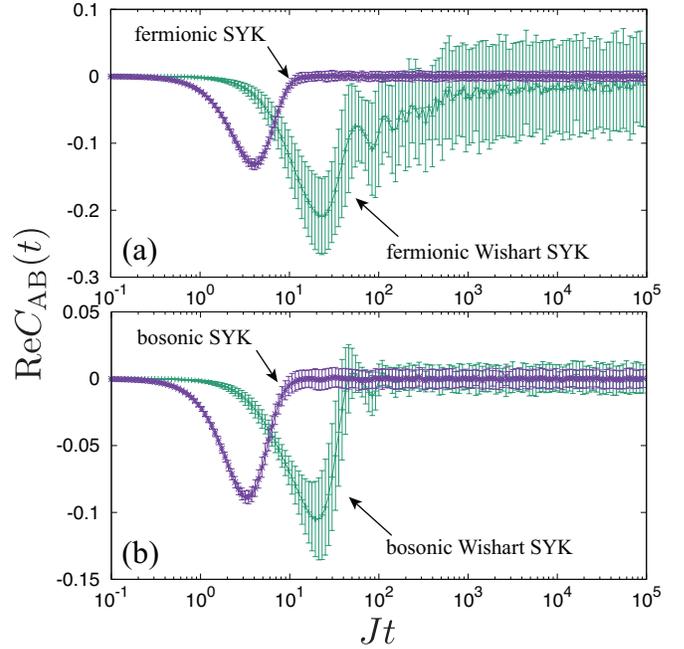}
\end{center}
\caption{
Time dependence of $\mathrm{Re} C_\mathrm{AB}(t)$ for $N=12$ and $N_{\mathrm{P}}=6$ when the initial state is an energy eigenstate.
The eigenenergies of the initial states nearly equal $-0.1J$ for $\hat{H}_\mathrm{SYK}$ and $0.1J$ for $\hat{H}_\mathrm{wSYK}$.
The number of samples is $128$.
}
\label{fig:OTOC_timedep}
\end{figure}

Figure~\ref{fig:OTOC_timedep} shows time dependence of OTOC for the original SYK model and the Wishart SYK model. We set $A=c_1$ ($b_1$) and $B=c_1^\dag$ ($b_1^\dag$) for the fermionic (bosonic) case. 
Figure~\ref{fig:OTOC_timedep} (a) shows the case of the fermionic models.
While the OTOC for the original SYK model shows quick relaxation, the OTOC for the Wishart SYK model shows slower relaxation with large temporal fluctuations at late times. We note that the temporal fluctuations of the tripartite mutual information of the clean SYK model are also larger than that of the disordered model~\cite{Iyoda2018PhysRevA.97.042330}.
On the other hand, 
Fig.~\ref{fig:OTOC_timedep} (b) shows that the temporal fluctuations of the OTOCs of the bosonic models are smaller than that of the fermionic Wishart SYK model at late times.

To investigate the dynamics of the OTOCs more systematically, we consider the effective dimension of the initial state.
We write the initial state as
\begin{align}
|\Psi\rangle=\sum_i \sum_{\alpha=1}^{d_i} c_{i}^\alpha \ket{E_{i}^\alpha},
\end{align}
where $\ket{E_i^\alpha}$ is an eigenstate with eigenenergy $E_i$ with $\alpha$ being a label of degeneracies, and $d_i$ represents the degeneracy of $E_i$.
The effective dimension of $|\Psi \rangle$ is defined as
\begin{align}
D_\mathrm{eff}(\ket{\Psi}):=\left(\sum_i p_i^2\right)^{-1},
\end{align}
where $p_i:=\sum_{\alpha=1}^{d_i} |c_i^\alpha|^2$. The effective dimension has been investigated in the context of relaxation of the expectation value of an observable at late times~\cite{Reimann2008PhysRevLett.101.190403,Short20121367-2630-14-1-013063}. 
The temporal fluctuations of an observable $O$ around its long time average, written as $\Delta O$, are bounded as $(\Delta O)^2 \leq CD_G/D_{\mathrm{eff}}$, where $C$ is a constant independent of the system size and $D_G$ is the maximum degeneracy of energy gaps.
If the temporal fluctuation is small, the expectation value nearly equals the long time average in almost all times after relaxation time.
Thus, a large effective dimension implies relaxation of the expectation value to a stationary value.
Although the OTOC cannot be written as the expectation value of any single observable, we expect that a similar bound holds for the temporal fluctuations of the OTOC.
\begin{figure}[t]
\begin{center}
\includegraphics[width=\linewidth]{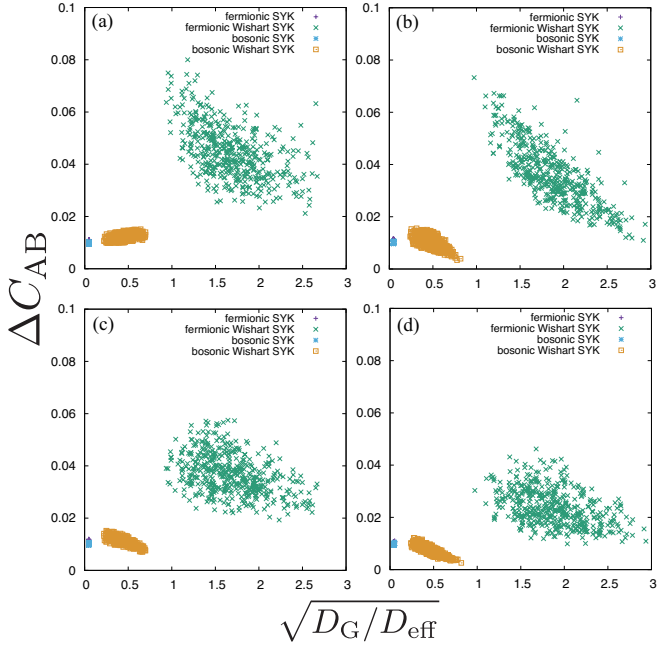}
\end{center}
\caption{
Temporal fluctuations of OTOC $\Delta C_\mathrm{AB}$ and $\sqrt{D_G/D_\mathrm{eff}}$ 
for $N=12$, $N_{\mathrm{P}}=6$ with a single disorder realization.
The pairs of observables $(A,B)$ are 
(a) $(a^\dag_1,a^\dag_1)$, 
(b) $(a^\dag_1,a_1)$, 
(c) $(a_1,a^\dag_1)$, and 
(d) $(a_1,a_1)$,
where $a_1=c_1$ or $b_1$ respectively for the fermionic or bosonic model. 
The time average is taken with $JT=10^4$.
}
\label{fig:OTOC_Deff}
\end{figure}

Figure~\ref{fig:OTOC_Deff} is a scatter plot between $\Delta C_{\mathrm{AB}}$ and $\sqrt{D_G/D_\mathrm{eff}}$. Each point represents a state in the computational basis.
We denote by $\ket{0}$ ($\ket{1}$) the empty (occupied) state at each site, and adopt the product states from $\ket{000\cdots 111}$ to $\ket{111\cdots 000}$ as the computational basis states.
For some pairs of observables, there are states whose OTOC and temporal fluctuations are trivially zero. We omit such trivial results from Fig.~\ref{fig:OTOC_Deff}. 
While the SYK model has small $\sqrt{D_G/D_\mathrm{eff}}$ and small temporal fluctuations, $\sqrt{D_G/D_\mathrm{eff}}$ and $\Delta C_{\mathrm{AB}}$ of the Wishart SYK models tend to be larger.
Thus, we expect that some relationship like $(\Delta O)^2 \leq CD_G/D_{\mathrm{eff}}$ can be valid for the case of OTOC.
It would be an interesting challenge for our future investigations to prove this rigorously.
We note that the effective dimensions of the fermionic/bosonic Wishart SYK models are of the same order of magnitude.

With regard to the OTOC, the bosonic/fermionic SYK models are qualitatively similar, which is consistent with Ref.~\cite{Fu2016PhysRevB.94.035135}.
However, we remark that the bosonic SYK model exhibits the glassy behavior, which is absent in the fermionic SYK model~\cite{Fu2016PhysRevB.94.035135}.

\section{Level statistics}
\label{sec:level}

In this section, we consider the level statistics of the Hamiltonian of the SYK models.
The level statistics has been well investigated to diagnose the conventional quantum chaos in quantum many-body systems~\cite{Stockmann1999,MehtaTextBook}.  It is known that the distribution of energy level spacings follows the Poisson distribution if the system is integrable and the Wigner-Dyson distribution if the system is non-integrable. The Wigner-Dyson distribution is classified into three classes GOE, GUE, and GSE corresponding to symmetries of the Hamiltonian. The level statistics of the (SUSY) SYK model is investigated in Refs.~\cite{You2017PhysRevB.95.115150,Kanazawa2017JHEP50}. 

We adopt the ratio of consecutive level spacings~\cite{Atas2013PhysRevLett.110.084101} to examine the level statistics of the fermionic/bosonic (Wishart) SYK models.
We assume that $E_i<E_j$ ($i<j$).
We define the nearest-neighbor spacing as $S_i:=E_{i+1}-E_i$. The ratio of consecutive level spacings is then defined as
\begin{align}
\tilde{r}_i
&:=
\frac{\min(S_i,S_{i-1})}{\max(S_i,S_{i-1})}
=
\min(r_i, 1/r_i),
\\
r_i
&:=
S_i/{S_{i-1}}.
\end{align}
By definition, $\tilde{r}_i$ takes a value within $0\leq \tilde{r}_i\leq 1$.

The level statistics of the energy level spacings is described by the Wigner-Dyson (Poisson) distribution for the non-integrable (integrable) systems, respectively.
As shown in Ref.~\cite{Atas2013PhysRevLett.110.084101}, the corresponding forms of $P(\tilde{r})$ are given by
\begin{align}
P_{\mathrm{Poisson}}(\tilde{r})
&:=
2/(1+\tilde{r})^2,
\\
P^\beta_{\mathrm{Wigner}}(\tilde{r})
&:=
\frac{2}{Z_\beta}
\frac{(\tilde{r}+\tilde{r}^2)^\beta}{(1+\tilde{r}+\tilde{r}^2)^{1+(3/2)\beta}},
\end{align}
where $\beta=1$ (GOE), $2$ (GUE), $4$ (GSE),
$Z_1=8/27$, $Z_2=(4/81)(\pi/\sqrt{3})$, and $Z_4=(4/729)(\pi/\sqrt{3})$.

\begin{figure}[t]
\begin{center}
\includegraphics[width=\linewidth]{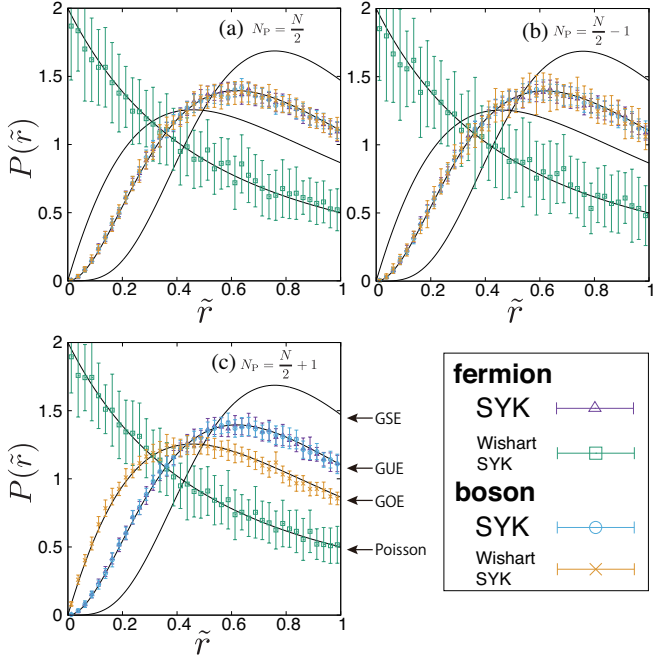}
\end{center}
\caption{
The distribution of the ratio of consecutive level spacings $P(\tilde{r})$ of $H_\mathrm{SYK}$ and $H_\mathrm{wSYK}$ for different $N_{\mathrm{P}}$'s.
The number of sites is $N=16$, and the number of samples is $24$.
}
\label{fig:Level_statistics}
\end{figure}

Figure~\ref{fig:Level_statistics} shows the distribution of the ratio of consecutive level spacings $P(\tilde{r})$ of the SYK models. The average is taken over $24$ samples of disorder, and the error bars represent the standard deviation. 
As shown in Fig.~\ref{fig:Level_statistics}, the distributions of the fermionic/bosonic SYK models are in good agreement with the GUE prediction. 
For the bosonic Wishart SYK model, the distribution follows that of the GUE, except for the case of $N_{\mathrm{P}}=N/2+1$ where it follows the GOE distribution (see Fig.~\ref{fig:Level_statistics}(c)). 
As will be shown in Appendix~\ref{sec:AppAntiUnitary}, we can understand the origin of the GOE distribution for this special case by constructing an anti-unitary operator which commutes with the Hamiltonian. Such an operator can be constructed only when $N_{\mathrm{P}}=N/2+1$ (see Appendix \ref{sec:AppAntiUnitary}). 

It is noteworthy that the distribution of the fermionic Wishart SYK model matches the Poisson distribution, implying two possibilities. One is that the fermionic Wishart SYK model is integrable, and the other is that we missed another symmetry of the Hamiltonian (though we have already considered the $U(1)$ symmetry corresponding to the particle number conservation). In the next section, we will show that the fermionic Wishart SYK model is, in fact, integrable.

\begin{figure}[t]
\begin{center}
\includegraphics[width=0.8\linewidth]{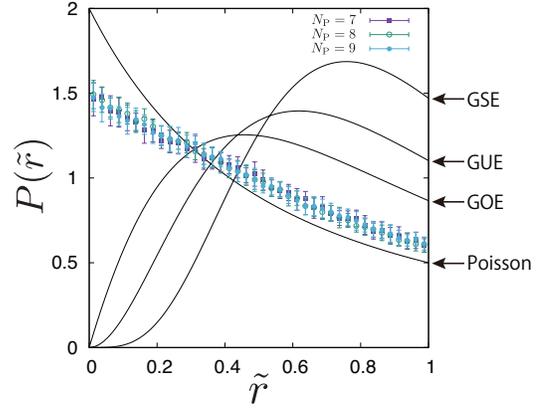}
\end{center}
\caption{
The distribution of the ratio of consecutive level spacings $P(\tilde{r})$ of $H_\mathrm{bSYK, 2body}$.
The number of sites is $N=16$, and the number of samples is $24$.
}
\label{fig:Level_statistics2}
\end{figure}

We also show the level statistics of the bosonic SYK model with 2-body interactions, whose Hamiltonian is defined as
\begin{align}
\label{eq:bSYK2}
H_\mathrm{bSYK, 2body}
:=
\frac{1}{N}
\sum_{1\leq i\leq N \atop 1\leq j\leq N}
J_{i;j}
b_i^\dag 
b_j,
\end{align}
where $J_{i;j}$ is sampled from the complex Gaussian distribution, satisfying $J_{i;j}=J_{j,i}^*$.
Figure~\ref{fig:Level_statistics2} shows that the level statistics is closest to the Poisson distribution.
This result can be understood by relating the bosonic SYK model with 2-body interactions to the fermionic Wishart SYK model with 4-body interactions.
The fermionic Wishart SYK model is mapped to Eq.~(\ref{eq:RGmodel}) in the next section. 
By identifying a two-fermion pairing term with a hard-core boson operator, we find that the Hamiltonian (\ref{eq:RGmodel}) is very similar to Eq.~(\ref{eq:bSYK2}). The only difference between Eqs.~(\ref{eq:RGmodel}) and (\ref{eq:bSYK2}) is the distribution from which the coupling strength is sampled, which is not relevant to the integrability of the fermionic Wishart SYK model.

\section{Integrability of the fermionic Wishart SYK model}
\label{sec:integrability}
In this section, we show the integrability of the fermionic Wishart SYK model by mapping it to the Richardson-Gaudin model~\cite{Richardson1965doi:10.1063/1.1704367,gaudin_2014}, which is known to be integrable by the algebraic Bethe ansatz~\cite{PAN19981PhysLettB, PanFeng19980305-4470-31-32-009, PAN1999120AnnPhys, Balantekin2007PhysRevC.75.064304,Samaj_bajnok_2013}.
We also explicitly construct the mutually commuting conserved quantities.
We examine the degenerate structure in the excited states shown in Fig.~\ref{fig:Deg_wSYK}(a).

\subsection{Mapping to the Richardson-Gaudin model}
\label{sec:integrabilityMap}
For simplicity, we first assume that $N$ is even with $N = 2M$ ($M\in \mathbb{N}$) and the coupling strengths $J_{k,l}$ are real. It is well known that any real skew-symmetric matrix $J$ can be brought into the following canonical form:
\begin{align}
\label{eq:canonical_form}
O^TJO
=
\left(
\begin{matrix}
0 & \lambda_1 &&&&&\\
-\lambda_1 & 0 &&&&&\\
&&0 & \lambda_2 &&&\\
&&-\lambda_2 & 0&&&\\
&&&&\ddots && \\
&&&&&0 & \lambda_M \\
&&&&&-\lambda_M & 0
\end{matrix}
\right),
\end{align}
where the matrix $O$ is orthogonal, $\lambda_j\in\mathbb{R}$ ($j=1,2,\cdots,M$) and zero elements of $J$ are left empty. 
In a generic case, $\lambda_j> 0$ is expected.
With the matrix $O$, we introduce a new set of fermionic operators
\begin{align}
\label{eq:new_fermions}
(f_{1,\uparrow},f_{1,\downarrow},\cdots,f_{M,\uparrow},f_{M,\downarrow})
:=
(c_1,c_2,\cdots,c_{2M-1},c_{2M})O,
\end{align}
which satisfy the anti-commutation relations:
\begin{align}
\{f_{j,\sigma},f^\dag_{k,\tau}\}
&=\delta_{j,k}\delta_{\sigma,\tau},
\\
\{f_{j,\sigma},f_{k,\tau}\}
&=
\{f^\dag_{j,\sigma},f^\dag_{k,\tau}\}
=
0.
\end{align}
We also define the corresponding number operators: $n_{j,\sigma}:=f^\dag_{j,\sigma}f_{j,\sigma}$ and $n_j:=n_{j,\uparrow}+n_{j,\downarrow}$.
In terms of the new fermion operators, the operator $Q$ and the Hamiltonian are written as
\begin{align}
Q
&=
\sum_{j=1}^M
Q_j,
\\
\label{eq:RGmodel_Qj}
Q_j&:=
\lambda_j f_{j,\uparrow}f_{j,\downarrow},
\\
\label{eq:RGmodel}
H
&=
\left(
\sum_{j=1}^M \lambda_j f^\dag_{j,\downarrow}f^\dag_{j,\uparrow}
\right)
\left(
\sum_{k=1}^M \lambda_k f_{k,\uparrow}f_{k,\downarrow}
\right).
\end{align}
This is nothing but a particular case of the Richardson-Gaudin model~\cite{Richardson1965doi:10.1063/1.1704367,gaudin_2014,Samaj_bajnok_2013,Dickhoff_Neck}, which is known to be integrable.
We note that the above mapping is possible in spite of the disorder and the distribution of the disordered coupling strength is not important for the mapping.

In the following, using the new set of fermionic operators~(\ref{eq:new_fermions}), we explicitly show the integrability of the fermionic Wishart SYK model and the degeneracies in the excited states shown in Fig.~\ref{fig:Deg_wSYK} (a).

\subsection{Integrability of the fermionic Wishart SYK model}
\label{sec:integrabilityB}
We now show that the fermionic Wishart SYK model is integrable in the sense of Ref.~\cite{Caux2011_1742-5468-2011-02-P02023} by explicitly constructing the conserved quantities.
The Hamiltonian~(\ref{eq:RGmodel}) can be written as a sum of the mutually commuting operators $\{h_j\}$ in accordance with Ref.~\cite{Pehlivan2008arXiv0806.1810P}:
\begin{align}
H=&\sum_{j=1}^M \lambda_j^2 h_j,
\\
\nonumber
h_j:=&
\eta^+_j \eta^-_j
-
2
\sum_{k(\neq j)}
\frac{\lambda_k^2}{\lambda_k^2-\lambda_j^2}\eta^z_j\eta^z_k\\
&-
\sum_{k(\neq j)}
\frac{\lambda_k\lambda_j}{\lambda_k^2-\lambda_j^2}
\left(
\eta^+_j \eta^-_k+\eta^+_k \eta^-_j
\right), 
\label{eq:mutually_commuting_ops}
\end{align}
where we defined $\eta$ operators as
\begin{align}
&
\eta_j^+:=f^\dag_{j,\downarrow}f^\dag_{j,\uparrow}
,\quad
\eta_j^-:=f_{j,\uparrow}f_{j,\downarrow},
\\
&
\eta_j^z:=\frac{1}{2}(
f^\dag_{j,\uparrow}f_{j,\uparrow}
+f^\dag_{j,\downarrow}f_{j,\downarrow}
-1
).
\end{align}
One can verify that $h_j$'s mutually commute: $[h_j,h_k]=0$. In addition, if these commuting operators are algebraically independent, the Hamiltonian is integrable. We show their linear independence in Appendix~\ref{sec:AppIndependence} and expect that their algebraic independence also holds.
From the above properties of the Hamiltonian, we expect that the Hamiltonian~(\ref{eq:RGmodel}) is $O(M)$ {\it quantum integrable} according to \cite{Caux2011_1742-5468-2011-02-P02023}.

We also note that the integrability of the fermionic Wishart SYK model can be shown by the algebraic Bethe ansatz (ABA).
In Appendices~\ref{sec:AppNotation} and \ref{sec:AppABA}, we review the previous results of the ABA in our context.
In Appendix~\ref{sec:AppNotation}, we define the generators of SU(2) and another algebra, and in Appendix~\ref{sec:AppABA}, we write down the ansatz states in the ABA. 

\subsection{Degeneracy of energy eigenstates}
We next investigate the degeneracy of the ground states and the excited states.
We can easily construct some of the ground states, which take the form of
\begin{align}
\label{eq:state_zes}
(f^\dag_{1,\sigma_1})^{n_1}
(f^\dag_{2,\sigma_2})^{n_2}
\cdots
(f^\dag_{M,\sigma_M})^{n_M}
|\mathrm{vac}\rangle,
\end{align}
where $\sigma_j=\uparrow$ or $\downarrow$,
$n_j=0$ or $1$,
and 
$|\mathrm{vac}\rangle$ is the vacuum state annihilated by $f_{j,\sigma}$ for all $(j,\sigma)$.
One can verify that these states are annihilated by $Q=\sum_{j=1}^M \lambda_j f_{j,\uparrow}f_{j,\downarrow}$ and thus by $H=Q^\dag Q$.
Therefore, these states are zero-energy states of the Hamiltonian.
The number of them amounts to $3^M=3^{N/2}$, which is less than the total number of the zero-energy states $Z_N$.
We show how to find the rest of the zero-energy states in the following. 

As shown in Fig.~\ref{fig:Deg_wSYK} (a), there are many degenerate excited states, whose degeneracy is given by $2^l$ with $l=0,2,4$, and $6$ for $(N,N_{\mathrm{P}})=(16,8)$. This can be understood by considering the block structure of the canonical form of the coupling strength (\ref{eq:canonical_form}).
We will explain the structure of the degeneracy for the case of $N_{\mathrm{P}}=N/2$ as an example.
Let us consider a state defined as
\begin{align}
f^\dag_{1,\sigma_1}
f^\dag_{2,\sigma_2}
\cdots
f^\dag_{M,\sigma_M}
|\mathrm{vac}\rangle,
\end{align}
which is a special case of Eq.~(\ref{eq:state_zes}) with $n_i=1$ ($i=1,\cdots,M$).
We also consider the following states which are a little different from the above state:
\begin{align}
&
f^\dag_{1,\sigma_1}
f^\dag_{2,\sigma_2}
\cdots
f^\dag_{M-2,\sigma_{M-2}}
f^\dag_{M-1,\uparrow}
f^\dag_{M-1,\downarrow}
|\mathrm{vac}\rangle,
\\
&
f^\dag_{1,\sigma_1}
f^\dag_{2,\sigma_2}
\cdots
f^\dag_{M-2,\sigma_{M-2}}
f^\dag_{M,\uparrow}
f^\dag_{M,\downarrow}
|\mathrm{vac}\rangle.
\end{align}
These states are annihilated by $Q_j$ ($j=1,\cdots,M-2$) in Eq.~(\ref{eq:RGmodel_Qj}), which results in $2^{M-2}$-fold degeneracy.
Thus, in order to obtain the eigenenergies of the Hamiltonian in the sector of $n_i=1$ ($i=1,\cdots,M-2$) and $n_j\neq 1$ ($j=M-1,M$), it is enough to consider the following restricted Hamiltonian:
\begin{align}
\label{eq:rest_Ham}
H_{M-1,M}:=
\left(
\sum_{j=M-1}^M \lambda_j f^\dag_{j,\downarrow}f^\dag_{j,\uparrow}
\right)
\left(
\sum_{k=M-1}^M \lambda_k f_{k,\uparrow}f_{k,\downarrow}
\right).
\end{align}
This Hamiltonian is represented by a $2\times 2$ matrix, and the eigenenergies are given by $0$ and $\lambda_{M-1}^2+\lambda_M^2(>0)$.
By defining $H_{i,j}$ in the same manner, the same discussion applies to $H_{i,j}$ for any $i$ and $j$ $(i\neq j)$.
These excited eigenstates are degenerate and their degeneracy is $2^{M-2}$, where we assume that $\lambda_i^2+\lambda_j^2\neq \lambda_k^2+\lambda_l^2$ for $(i,j)\neq(k,l)$.
The number of excited eigenenergies with $2^{M-2}$-fold degeneracy is easily obtained as $C(M,2)$.

We can also explain $2^{M-4}$-fold degeneracy by considering the Hamiltonian in the sector of $n_i=1$ ($i\neq i_1,i_2,i_3,i_4$) and $n_i\neq 1$ ($i=i_1,i_2,i_3,i_4$).
The restricted Hamiltonian is defined as
\begin{align}
H_\mathrm{I}:=
\left(
\sum_{j\in\mathrm{I}} \lambda_j f^\dag_{j,\downarrow}f^\dag_{j,\uparrow}
\right)
\left(
\sum_{k\in\mathrm{I}} \lambda_k f_{k,\uparrow}f_{k,\downarrow}
\right),
\end{align}
where $\mathrm{I}:=\{i_1,i_2,i_3,i_4\}$.
This Hamiltonian is represented by a $6\times 6$ matrix ($C(m,m/2)=6$ with $m=|I|=4$, where $|\cdot|$ is the number of elements).
There are two eigenstates with zero eigenenergies, which is understood similarly as the discussion about the number of zero-energy eigenstates in Sec.~\ref{sec:model}~\ref{sec:modelZES}.
In this case, the number of zero-energy eigenstates of $H_\mathrm{I}$ is counted as $C(4,2)-C(4,1)=2$.
Thus, the number of excited eigenenergies with $2^{M-4}$-fold degeneracy is calculated as $C(M,4)\{C(4,2)-(C(4,2)-C(4,1))\}=C(M,4)C(4,1)$.
Similarly, we find that the number of excited eigenenergies with $2^{M-2l}$-fold degeneracy is $C(M,2l)C(2l,l-1)$.

We note that the maximum degeneracy is brought by the smallest restricted Hamiltonians $H_{i,j}$.
Naturally, we expect that the spectrum is broader if $|I|$ is larger.
Thus, the above result explains that the degeneracy becomes smaller in the higher energy region in Fig.~\ref{fig:Deg_wSYK} (a).

For general $N$ and $N_{\mathrm{P}}$, the above discussion applies in the same manner.
We briefly comment on the case of odd $N$.
When $N$ is odd, the canonical form of the coupling strengths becomes
\begin{align}
\label{eq:canonical_form2}
O^TJO
=
\left(
\begin{matrix}
0 & \lambda_1 &&&&&&\\
-\lambda_1 & 0 &&&&&&\\
&&0 & \lambda_2 &&&&\\
&&-\lambda_2 & 0&&&&\\
&&&&\ddots && &\\
&&&&&0 & \lambda_M &\\
&&&&&-\lambda_M & 0&\\
&&&&&&&0
\end{matrix}
\right),
\end{align}
where the last block is $1\times 1$ and its element is $0$.
With this structure, the Hamiltonian splits into two parts corresponding to the fermion number of the last block.
When the number of the fermion in the last block is $1$ ($0$), the structure of degenerate excited states is the same as one with $N-1$ sites and $N_{\mathrm{P}}-1$ particles ($N-1$ sites and $N_{\mathrm{P}}$ particles).

We also note that the excited states can be generated algebraically for each restricted Hamiltonian.
We define lowering operators as
\begin{align}
&
\cS^-_m
:=
\sum_{j=1}^M(\lambda_j)^{2m+1}S_j^-,\quad
S_j^-:=f^\dag_{j,\downarrow}f_{j,\uparrow}.
\end{align}
We denote the restricted Hamiltonian by $\hat{H}_\mathrm{I}\otimes \hat{1}_\mathrm{\overline{I}}$, where $\overline{\mathrm{I}}$ is the complement of $\mathrm{I}$ and $\hat{1}_\mathrm{\overline{I}}$ is the identity operator defined on $\overline{\mathrm{I}}$.
One of the energy eigenstates of the Hamiltonian can be written by $\ket{\Phi,\mathrm{I}}\otimes\ket{\Psi,\overline{\mathrm{I}}}$, where $\ket{\Phi,\mathrm{I}}$ is an excited eigenstate of $\hat{H}_\mathrm{I}$ and $\ket{\Psi,\overline{\mathrm{I}}}$ is the ``ferromagnetic" state defined by 
\begin{align}
\ket{\Psi,\overline{\mathrm{I}}}:=
\left(\prod_{j\in \mathrm{\overline{I}}}f_{j,\uparrow}^\dag\right)
|\mathrm{vac}_{\overline{\mathrm{I}}}\rangle,
\end{align}
where $|\mathrm{vac}_{\overline{\mathrm{I}}}\rangle$ is the vacuum state of $\overline{\mathrm{I}}$.
Acting with the lowering operators on $\ket{\Phi,\mathrm{I}}\otimes\ket{\Psi,\overline{\mathrm{I}}}$ repeatedly, we obtain the degenerate excited energy eigenstates.

\section{Conclusion}
\label{sec:conclusion}
In this paper, we have introduced a variant of the SYK model, which is referred to as the Wishart SYK model.
We have numerically investigated the energy spectrum of the original and the Wishart SYK models for fermions/bosons.
We have shown that there is a huge number of degeneracy in the ground state of the Wishart SYK models, and the degeneracy is given by Eq.~(\ref{eq:NumZES}). 
Then we have shown that the OTOC of the fermionic Wishart SYK model exhibits large temporal fluctuations at late times, i.e., 
$|J_\mathrm{mean}|t\gg 1$ or $|J|t\gg 1$.
The large fluctuations are explained by the small effective dimension brought by the huge degeneracy.
We have also numerically investigated the level statistics and found that the level statistics of the fermionic Wishart SYK model follows the Poisson distribution. 
Correspondingly, we have shown that the fermionic Wishart SYK model is integrable by mapping it onto a particular case of the Richardson-Gaudin model and by writing the Hamiltonian as a sum of mutually commuting operators.

Although the fermionic Wishart SYK model does not reproduce the key characteristics of the original SYK model such as maximally chaotic behavior, we believe that the model serves as a reference for assessing the effect of disorder on the original SYK model. We also hope that our results will shed some light on the disorder-free tensor models, which exhibit huge degeneracies in the spectrum~\cite{Krishnan2017JHEP1}, just as in the fermionic Wishart SYK model.

\begin{acknowledgments}
H.K. is grateful to Hajime Moriya for valuable comments.
E.I. and T.S. are supported by JSPS KAKENHI Grant Number JP16H02211.
E.I. is also supported by JSPS KAKENHI Grant Number JP15K20944.
H.K. is supported by JSPS KAKENHI Grant Number JP18K03445 and JP18H04478.
\end{acknowledgments}

\appendix
\section{Condition of the equalities in (\ref{eq:NumZES0}) and (\ref{eq:NumZES})}
\label{sec:AppEqualities}
In this appendix, we show that the equalities in (\ref{eq:NumZES0}) and (\ref{eq:NumZES}) indeed hold for the fermionic Wishart SYK model. 
The estimation of the zero-energy states (\ref{eq:NumZES0}) and (\ref{eq:NumZES}) is based on the rank-nullity theorem (\ref{eq:RankNullity}).
If the operator $Q$ is surjective, the equalities in (\ref{eq:NumZES0}) and (\ref{eq:NumZES}) are achieved.

We transform the operator $Q$ by the conjugation argument in Refs.~\cite{WITTEN1982253NPB,Hagendorf2013JSP150}.
Let us consider an invertible transformation $V$ and introduce the conjugated operator $\tilde{Q}$ as
\begin{align}
\tilde{Q}&:=VQV^{-1},
\\
V&:=\prod_{j\sigma} V_{j\sigma},
\\
V_{j\sigma}&:=1+(\sqrt{\lambda_j}-1)n_{j\sigma}.
\end{align}
We can easily check that the inverse operator of $V$ is given by
\begin{align}
V^{-1}&:=\prod_{j\sigma} V^{-1}_{j\sigma},
\\
V^{-1}_{j\sigma}&:=1+\{(\sqrt{\lambda_j})^{-1}-1\}n_{j\sigma}.
\end{align}
With the conjugated operator, we define the corresponding Hamiltonian as
\begin{align}
\tilde{H}_\mathrm{wSYK}:=\tilde{Q}^\dag\tilde{Q},
\end{align}
which has the same number of zero-energy eigenstates as $H_\mathrm{wSYK}$.
We note that the conjugated Hamiltonian $\tilde{H}_\mathrm{wSYK}$ corresponds to the quasispin limit discussed in Ref.~\cite{Balantekin2007PhysRevC.75.064304}.

By the above transformation, we find
\begin{align}
\tilde{Q}=\sum_{j=1}^M \eta^-_j.
\end{align}
We can identify $\eta^z_j$ and $\eta^\pm_j$ with angular momentum operators of an SU(2) spin.
Then, the conjugated operator $\tilde{Q}$ is the lowering operator of the total angular momentum of $M$ spins.
We define 
\begin{align}
\tilde{Q}^z=\sum_{j=1}^M \eta^z_j,
\end{align}
whose eigenvalues are written as $k/2$ with $k=-M, -M+1,\cdots,M-1,M$.
From the theory of angular momentum coupling, 
$\tilde{Q}$ is surjective in the subspaces spanned by the eigenstates with non-positive $k$, if $N_{\mathrm{P}}\leq \lfloor \frac{N+1}{2}\rfloor$.
Thus, the equalities in (\ref{eq:NumZES0}) and (\ref{eq:NumZES}) are achieved for the fermionic Wishart SYK model.

\section{Construction of an anti-unitary operator}
\label{sec:AppAntiUnitary}
In Sec.~\ref{sec:level}, we showed that the level statistics of the bosonic Wishart SYK model is GUE with $N_{\mathrm{P}}\neq N/2+1$ and GOE with $N_{\mathrm{P}}=N/2+1$. This result suggests that there exists an anti-unitary operator commuting with the Hamiltonian, which can be defined only for $N_{\mathrm{P}}=N/2+1$. In this appendix, we explicitly construct such an anti-unitary operator.

We first recall that from the random matrix theory, the level statistics is GOE (GSE) if $P^2=+I$ ($-I$) is satisfied, where $P$ is an anti-unitary operator commuting with the Hamiltonian.

We consider the Hamiltonian of the Wishart SYK model in the form of Eq.~(\ref{eq:Ham_wSYK}).
Let $P$ be a particle-hole operator defined as 
\begin{align}
P:=K\prod_{i=1}^N(a_i+a_i^\dag),
\end{align}
where $K$ is an operator of complex conjugation and $a_i=c_i$ ($b_i$) for the fermionic (bosonic) model.
We note that $P$ is anti-unitary.
For hard-core bosons, $P$ satisfies $P^2=+I$.
For fermions, $P$ satisfies $P^2=+I$ if $N=0~(\mathrm{mod}~4)$ and $P^2=-I$ if $N=2~(\mathrm{mod}~4)$.
We find that $PQP=Q^\dag$, $PQ^\dag P=Q$ and $PHP=QQ^\dag$ by simple calculations.

We define 
\begin{align}
\tilde{P}:=PQ|Q|^{-1},
\end{align}
where $|Q|:=\sqrt{Q^\dag Q}$.
We should note that $\tilde{P}$ can be defined only for $N_{\mathrm{P}}=N/2+1$, because $|Q|^{-1}$ cannot be defined in the other cases.
By simple calculations, we find that $\tilde{P}$ is anti-unitary and satisfies $\tilde{P}^2=\pm I$, and $[H,\tilde{P}]=[Q^\dag Q,\tilde{P}]=0$.
Thus, the existence of the operator $\tilde{P}$ explains why the level statistics of the bosonic Wishart SYK model obeys GOE only when $N_{\mathrm{P}}=N/2+1$ as shown in Fig.~\ref{fig:Level_statistics}.

Although this discussion holds both for the fermionic and bosonic Wishart SYK models, the level statistics of the fermionic Wishart SYK model is not GOE nor GSE, because it is integrable as discussed in Sec.~\ref{sec:integrability}.

As a side remark, let us also consider a variant of the fermionic Wishart SYK model
\begin{align}
\label{eq:Ham_wSYK8}
H_{\mathrm{wSYK},8}
&:=
Q_4^\dag Q_4,
\\
Q_4
&:=
\frac{1}{N}\sum_{1\leq k<l<m<n\leq N} J_{k,l,m,n}c_k c_l c_m c_n.
\end{align}
With this model, the above discussion of the construction of $\tilde{P}$ applies for $N_{\mathrm{P}}=N/2+2$, and we have numerically confirmed that the level statistics of the model is GSE for $N=14$ and GOE for $N=16$ (data not shown).
Thus, the model is unlikely to be integrable.

\section{Linear independence of the conserved charges}
\label{sec:AppIndependence}
We show the linear independence of the mutually commuting operators $\{h_j\}$ defined in (\ref{eq:mutually_commuting_ops}) of Sec.~\ref{sec:integrability} \ref{sec:integrabilityB}.
We first define the projection operator $P_j$ as
\begin{align}
P_j:=\eta^+_j \eta^-_j.
\end{align}
We also define the inner product of operators $A,B$ using the trace as
\begin{align}
(A,B):=\mathrm{tr}\left[A^\dag B\right]/2^{2M}.
\end{align}
The inner product of $P_j$ and $h_k$ is calculated as
\begin{align}
(P_j,h_k)
= 
\begin{cases}
1/4 & (j=k), \\
1/16 & (j\neq k).
\end{cases}
\end{align}

Let us consider the following equation:
\begin{align}
\label{eq:linear_comb}
\sum_{i=1}^M a_i h_i=0,
\end{align}
where $a_i\in\mathbb{C}$.
Taking the inner product of $P_j$ and the left-hand side of Eq.~(\ref{eq:linear_comb}), we obtain
\begin{align}
\label{eq:algebraic_indep}
L
\left(
\begin{matrix}
a_1 \\
a_2 \\
a_3 \\
\vdots\\
a_M
\end{matrix}
\right)=0,
\end{align}
where $L$ is an $M\times M$ matrix, whose elements are given by
\begin{align}
L_{jk}
=
\begin{cases}
1/4 & (j=k), \\
1/16 & (j\neq k).
\end{cases}
\end{align}
Since $\det L= (M+3)3^{M-1}/16^{M}\neq 0$, Eq.~(\ref{eq:algebraic_indep}) only has a solution $a_1=\cdots=a_M=0$, which implies the linear independence of $\{h_j\}$.

\section{Generator of SU(2) and another algebra}
\label{sec:AppNotation}
In this appendix, we define the generator of SU(2) and another algebra.
Let us first introduce the generators of SU(2) symmetries:
\begin{align}
&
S_j^+:=f^\dag_{j,\uparrow}f_{j,\downarrow},
\quad
S_j^-:=f^\dag_{j,\downarrow}f_{j,\uparrow},
\\
&
S_j^z:=\frac{1}{2}(f^\dag_{j,\uparrow}f_{j,\uparrow}
-f^\dag_{j,\downarrow}f_{j,\downarrow}
),
\\
&
\eta_j^+:=f^\dag_{j,\downarrow}f^\dag_{j,\uparrow},
\quad
\eta_j^-:=f_{j,\uparrow}f_{j,\downarrow},
\\
&
\eta_j^z:=\frac{1}{2}(
f^\dag_{j,\uparrow}f_{j,\uparrow}
+f^\dag_{j,\downarrow}f_{j,\downarrow}
-1
).
\end{align}
From the commutation relations among $f^\dag_{j,\sigma}$ and $f_{j,\sigma}$, we can verify the following relations:
\begin{align}
\label{eq:CR_Seta1}
&
[S_j^+,S_k^-]
=
2\delta_{j,k}S_j^z,
\quad
[S_j^z,S_k^\pm]
=
\pm\delta_{j,k}S_j^\pm,
\\
\label{eq:CR_Seta2}
&
[\eta_j^+,\eta_k^-]
=
2\delta_{j,k}\eta_j^z
,
\quad
[\eta_j^z,\eta_k^\pm]
=
\pm\delta_{j,k}\eta_j^\pm,
\\
\label{eq:CR_Seta3}
&
[S_j^\alpha,S_k^\alpha]
=
[\eta_j^\alpha,\eta_k^\alpha]
=
[\eta_j^\alpha,S_k^\beta]
=0,
\quad(\alpha,\beta=z,+,-).
\end{align}
We note that the SU(2) symmetry generated by $\eta$-pairing operators has been discussed in the context of the Hubbard model~\cite{Yang1989PhysRevLett.63.2144,Yang1990doi:10.1142/S0217984990000933}.
According to Refs.~\cite{PanFeng19980305-4470-31-32-009,PAN1999120AnnPhys}, we also introduce another algebra generated by
\begin{align}
&
\cS^\pm_m
:=
\sum_{j=1}^M(\lambda_j)^{2m+1}S_j^\pm,
\quad
\cS_m^z
:=
\sum_{j=1}^M(\lambda_j)^{2m}S_j^z,
\\
&
\cT^\pm_m
:=
\sum_{j=1}^M(\lambda_j)^{2m+1}\eta_j^\pm,
\quad
\cT_m^z
:=
\sum_{j=1}^M(\lambda_j)^{2m}\eta_j^z.
\end{align}
From the commutation relations from (\ref{eq:CR_Seta1}) to (\ref{eq:CR_Seta3}), we can verify the following relations:
\begin{align}
\label{eq:CR_ST}
&
[\cS_m^+,\cS_n^-]
=
2\cS_{m+n+1}^z,
\quad
[\cS_m^z,\cS_n^\pm]
=
\pm \cS_{m+n}^\pm,
\\
&
[\cT_m^+,\cT_n^-]
=
2\cT_{m+n+1}^z,
\quad
[\cT_m^z,\cT_n^\pm]
=
\pm \cT_{m+n}^\pm,
\\
&
[\cS_m^\alpha,\cS_n^\alpha]
=
[\cT_m^\alpha,\cT_n^\alpha]
=
[\cS_m^\alpha,\cT_n^\beta]
=0
\quad(\alpha,\beta=z,+,-).
\end{align}

\section{Fermionic Wishart SYK model with the algebraic Bethe ansatz}
\label{sec:AppABA}
We show the integrability of the fermionic Wishart SYK model with the algebraic Bethe ansatz~\cite{PAN19981PhysLettB, PanFeng19980305-4470-31-32-009, PAN1999120AnnPhys, Balantekin2007PhysRevC.75.064304,Samaj_bajnok_2013}.
With $f_{j,\sigma}$ and $f^\dag_{j,\sigma}$, we introduce $S_j^\alpha$ and $\eta_j^\alpha$ ($\alpha=+,-,z$) and $\cS_m^\alpha$ and $\cT_m^\alpha$ ($\alpha=+,-,z$) as in Appendix~\ref{sec:AppNotation}.
Using $\cS_m^\alpha$ and $\cT_m^\alpha$, we write the Hamiltonian of the fermionic Wishart SYK model as
\begin{align}
\label{eq:Ham_wSYK_ABA}
H=\mathcal{T}_0^+\mathcal{T}_0^-.
\end{align}
One can show that the Hamiltonian commutes with all of $\cS^\alpha_j$. Thus, we can find many conserved charges, for example, $\cC:=i(\cS_0^+-\cS_0^-)$.

The Hamiltonian (\ref{eq:Ham_wSYK_ABA}) can be diagonalized by using the algebraic Bethe ansatz (ABA).
To see this, let us introduce the following operators
\begin{align}
\cT^\pm(x)
&=
\sum_{j=1}^M
\frac{\lambda_j}{1-(\lambda_j)^2 x}
\eta_j^\pm,
\\
\cT^z(x)
&=
\sum_{j=1}^M
\frac{(\lambda_j)^2}{1-(\lambda_j)^2 x}
\eta_j^z,
\end{align}
where $\cT^\pm(0)=\cT^\pm_0$.
The key relations for the ABA are
\begin{align}
[\cT^+(x),\cT^-(0)]
&=
[\cT^+(0),\cT^-(x)]
=
2\cT^z(x),
\\
[\cT^z(x),\cT^\pm(y)]
&=
\pm
\frac{\cT^\pm(x)-\cT^\pm(y)}{x-y},
\\
\cT^-(x)
|\mathrm{vac}\rangle
&=
0,
\\
\cT^z(x)
|\mathrm{vac}\rangle
&=
-\frac{1}{2}
\left(
\sum_{j=1}^M
\frac{(\lambda_j)^2}{1-(\lambda_j)^2 x}
\right)
|\mathrm{vac}\rangle.
\end{align}

The eigenstates of $H$ can be constructed by acting with $\cT^+(z_j)$ $(j=1,2,\cdots,n)$ on the vacuum. The ansatz state reads
\begin{align}
\label{eq:ansatz_state}
|\Psi(z_1,z_2,\cdots,z_n)\rangle
:=
\cT^+(z_1)\cT^+(z_2)\cdots \cT^+(z_n)|\mathrm{vac}\rangle.
\end{align}
Here we assume that $z_j$'s are distinct.
By acting with $H$ on this state, we obtain
\begin{multline}
H|\Psi(z_1,z_2,\cdots,z_n)\rangle
=
\\
\sum_{j=1}^n
\left(
\sum_{l=1}^M\frac{(\lambda_l)^2}{1-(\lambda_l)^2 z_j}
+
2\sum_{k=1\atop k\neq j}^n
\frac{1}{z_j-z_k}
\right)
\\
|\Psi(0,z_1,\cdots,z_{j-1},z_{j+1},\cdots,z_n)\rangle.
\end{multline}
Suppose that none of $z_j$ $(j=1,2,\cdots,n)$ is $0$.
Then, the ansatz state~(\ref{eq:ansatz_state}) is an eigenstate of $H$ with eigenvalue $0$, if $z_j$'s satisfy the following Bethe equations:
\begin{align}
\sum_{l=1}^M
\frac{(\lambda_l)^2}{1-(\lambda_l)^2 z_j}
+
2\sum_{k=1 \atop k\neq j}^n
\frac{1}{z_j-z_k}=0
\end{align}
for all $j=1,2,\cdots,n$.
The situation is different when one of $z_j$'s is $0$.
In this case, for example, $|\Psi(0,z_2,\cdots,z_n)\rangle$ is an eigenstate of $H$ with eigenvalue
\begin{align}
E=\sum_{l=1}^M(\lambda_l)^2
-
\sum_{k=2}^n\frac{2}{z_k},
\end{align}
demanding that the following equations hold for all $j=2,3,\cdots,n$:
\begin{align}
\sum_{l=1}^M
\frac{(\lambda_l)^2}{1-(\lambda_l)^2 z_j}
+
\frac{2}{z_j}
+
2\sum_{k=2\atop k\neq j}^n
\frac{1}{z_j-z_k}=0.
\end{align}
This reproduces the previous results in \cite{Balantekin2007PhysRevC.75.064304}. 

\bibliographystyle{h-physrev}
\bibliography{Wishart}

\end{document}